# High-density and Secure Data Transmission via Linear Combinations

*Preliminary version*

Vince Grolmusz [*]

## Abstract

Suppose that there are $n$ Senders and $n$ Receivers. Our goal is to send long messages from Sender $i$ to Receiver $i$ such that no other receiver can retrieve the message intended for Receiver $i$. The task can easily be completed using $n$ private channels between the pairs. Solutions, using one channel needs either encryption or switching elements for routing the messages to their addressee.

The main result of the present work is a description of a network in which The Senders and the Receivers are connected with only $n^{o(1)}$ channels; the encoding and de-coding is nothing else just very fast linear combinations of the message-bits; and there are no switching or routing-elements in the network, just linear combinations are computed, with fixed connections (channels or wires).

In the proofs we do not use *any* unproven cryptographical or complexity theoretical assumptions.

## 1 Introduction

Suppose that there are given $n$ Senders $S_1, S_2, \ldots, S_n$ and $n$ Receivers $R_1, R_2, \ldots, R_n$. Our goal is to send long messages from $S_i$ to $R_i$, for $i = 1, 2, \ldots, n$ such that

(a) $R_i$ can easily retrieve the message of $S_i$, for $i = 1, 2, \ldots, n$

(b) $R_i$ cannot retrieve the message of $S_j$ for any $j \neq i$.

An obvious method for doing this is connecting $S_i$ with $R_i$ with private channels, that is, we need $n$ parallel channels for the $n$ Senders and the $n$ Receivers. The advantage of this solution is that $n$ bits can be sent in parallel, and the transmission is private, in the sense that $R_i$ receives only the transmission of $S_i$, for $i = 1, 2, \ldots, n$.

Another obvious solution is that all the Senders and Receivers use the same channel, and they transmit their messages one after the other, but in this case some sort of encryption should be used for the maintaining the privacy of the transmission.

The main result of the present work is a description of a network in which

(i) The Senders and the Receivers are connected with only $n^{o(1)}$ channels[1]

[*]Department of Computer Science, Eötvös University, Budapest, Pázmány P. stny. 1/C, H-1117 Budapest, Hungary; E-mail: grolmusz@cs.elte.hu

[1]Here $o(1)$ denotes a quantity which goes to 0 as $n$ goes to the infinity.

(ii) The encoding and de-coding is nothing else just linear combinations of the message-bits, and this linear combinations can be computed very fast,

(iii) There are no switching or routing-elements in the network, just linear combinations are computed, with fixed connections (channels or wires),

(iv) $R_i$ cannot learn much about any bit of the message of $S_j$ for any $j \neq i$, and can learn only a negligible amount of information on longer messages of $S_j$;

(v) In the proofs we do not use *any* unproven cryptographical or complexity theoretical or any other assumptions.

## 2 Preliminaries

### 2.1 The dot-product

We have defined the *alternative*, and the *0-a-strong* and the *1-a-strong* representations of polynomials in [2]. Note that the 0-a-strong representation, defined here, coincides with the a-strong representation of the paper [1].

Note also, that for prime or prime-power moduli, polynomials and their representations (defined below), coincide. That may be the reason that such definitions were not given prior to [2].

**Definition 1** *Let $m$ be a composite number $m = p_1^{e_1} p_2^{e_2} \cdots p_\ell^{e_\ell}$. Let $Z_m$ denote the ring of modulo $m$ integers. Let $f$ be a polynomial of $n$ variables over $Z_m$:*

$$f(x_1, x_2, \ldots, x_n) = \sum_{I \subset \{1,2,\ldots,n\}} a_I x_I,$$

*where $a_I \in Z_m$, $x_I = \prod_{i \in I} x_i$. Then we say that*

$$g(x_1, x_2, \ldots, x_n) = \sum_{I \subset \{1,2,\ldots,n\}} b_I x_I,$$

*is an*

- alternative representation *of $f$ modulo $m$, if*

$$\forall I \subset \{1, 2, \ldots, n\} \ \exists j \in \{1, 2, \ldots, \ell\}: \quad a_I \equiv b_I \pmod{p_j^{e_j}};$$

- 0-a-strong representation *of $f$ modulo $m$, if it is an alternative representation, and, furthermore, if for some $i$, $a_I \not\equiv b_I \pmod{p_i^{e_i}}$, then $b_I \equiv 0 \pmod{p_i^{e_i}}$;*

- 1-a-strong representation *of $f$ modulo $m$, if it is an alternative representation, and, furthermore, if for some $i$, $a_I \not\equiv b_I \pmod{p_i^{e_i}}$, then $a_I \equiv 0 \pmod{m}$;*

**Example 2** *Let $m = 6$, and let $f(x_1, x_2, x_3) = x_1 x_2 + x_2 x_3 + x_1 x_3$, then*

$$g(x_1, x_2, , x_3) = 3 x_1 x_2 + 4 x_2 x_3 + x_1 x_3$$

*is a 0-a-strong representation of f modulo 6;*

$$g(x_1, x_2, , x_3) = x_1x_2 + x_2x_3 + x_1x_3 + 3x_1^2 + 4x_2$$

*is a 1-a-strong representation of f modulo 6;*

$$g(x_1, x_2, , x_3) = 3x_1x_2 + 4x_2x_3 + x_1x_3 + 3x_1^2 + 4x_2$$

*is an alternative representation modulo 6.*

In other words, for modulus 6, in the alternative representation, each coefficient is correct either modulo 2 or modulo 3, but not necessarily both.

In the 0-a-strong representation, the 0 coefficients are always correct both modulo 2 and 3, the non-zeroes are allowed to be correct either modulo 2 or 3, and if they are not correct modulo one of them, say 2, then they should be 0 mod 2.

In the 1-a-strong representation, the non-zero coefficients of $f$ are correct for both moduli in $g$, but the zero coefficients of $f$ can be non-zero either modulo 2 or modulo 3 in $g$, but not both.

In [2] we proved the following theorem:

**Theorem 3** (i) *Let $m = p_1p_2$, where $p_1 \neq p_2$ are primes. Then a degree-2 1-a-strong representation of the dot-product*

$$f(x_1, x_2, \ldots, x_n, y_1, y_2, \ldots, y_n) = \sum_{i=1}^n x_i y_i$$

*can be computed with*

$$\exp(O(\sqrt{\log n \log \log n})) \qquad (2)$$

*multiplications.*

(ii) *Moreover, the representation of (i) can be computed on bilinear $\Sigma\Pi\Sigma$ circuits of size (2).*

In other words, we have shown, that instead of the usual dot-product

$$\sum_{i=1}^n x_i y_i$$

we can compute a polynomial of the form

$$\sum_{i=1}^n x_i y_i + 3g(x, y) + 4h(x, y) \qquad (3)$$

where both $g$ and $h$ has the following form:

$$\sum_{i \neq j} a_{ij} x_i y_j, \quad a_{ij} \bmod 6 \in \{0, 1\},$$

and no term $x_i y_j$ appears in both $f$ and $g$.

Moreover, by (ii), (3) can be computed as the sum:

$$\sum_{j=1}^t \left( \sum_{i=1}^n b_{ij} x_i \right) \left( \sum_{i=1}^n c_{ij} y_i \right) \qquad (4)$$

where $b_{ij}, c_{ij} \in \{0, 1\}$ and $t = \exp(O(\sqrt{\log n \log \log n})) = n^{o(1)}$.

## 3 The main idea

In this short preliminary report we are dealing with only the case, when the modulus is 6. Our results can easily be generalized for other non-prime-power composite moduli as well.

Polynomial (3) can be computed in form (4). Let us consider a fixed $x \in \{0,1\}^n$, and let us compute $z = (z_1, z_2, \ldots, z_t) \in \{0,1,2,3,4,5\}^t$, where

$$z_j = \sum_{i=1}^{n} b_{ij} x_i \bmod 6 \qquad (5)$$

that is, simply a linear combination, determined by (4), for $j = 1, 2, \ldots, t$. Or, in matrix notation, where $B = \{b_{ij}\}$ denotes an $n \times t$ matrix,

$$z = xB \bmod 6.$$

Certainly, different $x$'s may lead to the same $z$, since $t = n^{o(1)} << n$.

Now, from (4), the 1-a-strong representation of the dot-product of $x$ with any substitution of the $y$'s can be computed from the $t = n^{o(1)}$ values of $z$. In particular, plugging in

$$y^{(i)} = (0, 0, \ldots, \overset{i}{\overset{\frown}{1}}, 0, \ldots, 0),$$

we will get modulo 6 the following polynomial:

$$x_i + 4(x_{i_1} + x_{i_2} + \cdots + x_{i_\ell}) + 3(x_{j_1} + x_{j_2} + \cdots + x_{j_k}) \qquad (6)$$

where different indices denote different numbers.

Or, in other words, with the notation of $C = \{c_{ij}\}$ as an $n \times t$ matrix of $c_{ij}$ entries from (4),

$$x' = x + 4xU + 3xV = xBC^T = zC^T,$$

where $U$ and $V$ are some $n \times n$

## 4 Hyperdense transmission

We describe the transmission-protocol in rounds. In every round, every sender $S_i$ will transmit securely a bit $x_i$ to the corresponding receiver, $R_i$, $i = 1, 2, \ldots, n$. In $u$ consecutive rounds, every sender will send $u$ bits, that is, sending $u$-bit messages needs $u$ rounds of the following protocol.

A round is performed as follows:

Step 1 - Encoding - From the bits of $x = (x_1, x_2, \ldots, x_n)$ the mod 6 integers $z = (z_1, z_2, \ldots, z_t)$ is computed by linear combinations taken modulo 6:

$$z = xB \bmod 6,$$

Step 2 - Transmission - The mod 6 numbers $z_1, z_2, \ldots, z_t$ are sent on $t$ channels to the receivers.

Step 3 - Decoding - The linear transformation $x' = (x'_1, x'_2, \ldots, x'n) = zC^T$ is computed modulo 6 at the receivers' side, and number $x'_i$ is given to receiver $R_i$, $i = 1, 2, \ldots, n$. (Note, that because of the obvious information-theoretical reasons, generally it is not possible to retrieve bit $x_i$ from integer $x'_i$).

Step 4 - Pre-Filtering - A random $\pi : \{1, 2, \ldots, n\} \to \{1, 2, \ldots, n\}$ permutation is generated at the senders's side. Then for $j = 1, 2, \ldots, n$, if $x_{\pi(j)}$ is 1, then Steps 1, 2 and 3 are repeated for $x^{\pi(j)} \in \{0, 1\}^n$ instead of $x$, where $x^{\pi(j)}$ coincides with $x$, except on position $\pi(j)$, whereas $x^{\pi(j)}$ is 0. Let $x''_i$ denote the coordinate $i$ of $x_{\pi(j)}BC^T$. On the other hand, if $x_{\pi(j)}$ is 0, then nothing happens.

Step 5 - Post-Filtering - Now, receiver $R_i$ stores value $x'_i$ in its memory, and, additionally, sets register $r_i = x'_i$, and follows the following program after receiving any new $x''_i$, originating in Step 4:

if $x''_i = x'_i$ it does nothing;

if $x''_i = x'_i - 1$ then $R_i$ concludes that $x_i = 1$;

if $x''_i \equiv x'_i - 3$ then sets $r_i = r_i - 3 \bmod 6$;

if $x''_i \equiv x'_i - 4$ then sets $r_i = r_i - 4 \bmod 6$;

if $R_i$ never got in the round an $x''_i$ for which $x''_i = x'_i - 1$, then it concludes that $x_i = 0$.

After finishing the repetitions described in Step 4, the value of $x_i$ and the mod 6 value of $r_i$ should be the same; otherwise $R_i$ outputs: ERROR. This property can also be used for error-detection in the protocol.

**Theorem 4** *After performing one round, receiver $R_i$ retrieves the bit $x_i$, for $i = 1, 2, \ldots, n$.*

**Proof:** Clearly, $x'_i$ is equal to quantity (6); so decreasing any non-zero $x_j$ in the sum of (6) by 1, leads to either a decrease of 1 of the sum (in the case when exactly $x_i$ was decreased, or by 0 (when an $x_j$ was decreased with a 0 coefficient in (6)), or by 4 or 3 modulo 6, (when the coefficient of the decreased variable was 4 or 3, respectively). If we subtract from $x'_i$ all the $x_j$'s with coefficients 3 or 4, then $x_i$ will remain. □

## 4.1 An alternative filtering method

The following modification of the filtering steps of the protocol relies on the fact that if we one-by-one increase the value of $r$, then $3r$ will have period 2, and $4r$ will have period 3 modulo 6, but $r$ itself will have period 6, modulo 6.

So, we can modify Steps 4 and 5 as follows:

Step 4' - Pre-filtering-variant- A random $\pi : \{1, 2, \ldots, n\} \to \{1, 2, \ldots, n\}$ permutation is generated at the senders's side. Then for $j = 1, 2, \ldots, n$, if $x_{\pi(j)}$ is 1, then Steps 1, 2 and 3 are repeated for six values of $x^{\pi(j)} \in \{0, 1, 2, 3, 4, 5\}^n$ instead of $x$, where $x^{\pi(j)}$ coincides with $x$, except on position $\pi(j)$, whereas $x^{\pi(j)}$ is takes on values 0,1,2,3,4, and 5, one after the other. Let $x''_i$ denote the coordinate $i$ of $x_{\pi(j)}BC^T$. On the other hand, if $x_{\pi(j)}$ is 0, then nothing happens.

Step 5' - Post-Filtering-variant - Receiver $R_i$ stores value $x'_i$ in its memory, and follows the following program after receiving the 6 new $x''_i$'s in Step 4:

if if the period of $x''_i$ is 6 then $R_i$ concludes that $x_i = 1$;

if if the period of $x_i''$ is less than 6 then it does nothing

If it never realized in the turn that the period is 6, then $R_i$ concludes that $x_i = 0$.

Note, that this filtering method can be more applicable than the original in electrical-engineering applications.

## 5 The Security of the network

The security of the network-protocol relies on the independently generated random permutations $\pi$ in each round.

Let us review, what $R_i$ can learn from the bits, addressed to others. Clearly, $R_i$ will know the number of the 1-bits with coefficient 4 and also the number of the 1-bits with coefficient 3 in (6), but $R_i$ will not know the identity of the 1-bits.

**Acknowledgment.**
The author acknowledges the partial support of the István Széchenyi fellowship.